\documentclass[twocolumn,english,twocolumn,prl,amssymb,amsmath]{revtex4}
\usepackage[latin9]{inputenc}
\usepackage{graphicx,color}
\usepackage{amssymb}
\usepackage[bookmarks]{hyperref}
\usepackage{babel}

\newcommand{\beq}{\begin{equation}}
\newcommand{\eeq}{\end{equation}}
\newcommand{\bea}{\begin{eqnarray}}
\newcommand{\eea}{\end{eqnarray}}

\begin{document}

\title{Dynamical Quantum Phase Transitions in the Transverse Field Ising Model}

\author{M. Heyl}
\affiliation{Department of Physics, Arnold Sommerfeld Center for Theoretical Physics
and Center for NanoScience, Ludwig-Maximilians-Universit\"at M\"unchen,
Theresienstr.\ 37, 80333 Munich, Germany}
\affiliation{Institut f\"ur Theoretische Physik, Technische Universit\"at Dresden, 01062 Dresden, Germany}
\author{A. Polkovnikov}
\affiliation{Department of Physics, Boston University, 590 Commonwealth Ave., Boston, MA 02215, USA}
\author{S. Kehrein}
\affiliation{Department of Physics, Georg-August-Universit\"at G\"ottingen, Friedrich-Hund-Platz 1, 37077 G\"ottingen, Germany}

\begin{abstract}
A phase transition indicates a sudden change in the
properties of a large system. For temperature-driven phase transitions this is related to non-analytic behavior
of the free energy density at the critical temperature: The knowledge of the free energy density in one phase
is insufficient to predict the properties of the other phase. In this paper we show that a close analogue 
of this behavior can occur in the real time evolution of quantum systems, namely non-analytic behavior 
at a critical time. We denote such behavior a {\em dynamical phase transition} and explore its
properties in the transverse field Ising model. Specifically, we show that the equilibrium quantum phase transition
and the dynamical phase transition in this model are intimately related.

\end{abstract}

\maketitle

Phase transitions are one of the most remarkable phenomena occurring in many-particle systems. At a phase transition
a system undergoes a non-analytic change of its properties, for example the density at a temperature driven liquid-gas
transition, or the magnetization at a paramagnet-ferromagnet transition. What makes the theory of such equilibrium
phase transitions particularly fascinating is the observation that a perfectly well-behaved microscopic Hamiltonian without
any singular interactions can lead to non-analytic behavior in the thermodynamic limit of the many-particle system.
In fact, the occurrence of equilibrium phase transitions was initially a puzzling problem because one can easily
verify no go theorems for finite systems, therefore the thermodynamic limit is essential \cite{MFisher}. 

Today the theory of equilibrium phase transitions is well established, especially for classical systems undergoing
continuous transitions, where the powerful tool of renormalization theory bridges the gap from microscopic Hamiltonian
to universal macroscopic behavior. On the other hand, the behavior of non-equilibrium quantum many-body systems is by far 
less well understood. Recent experimental advances have triggered a lot of activity in this field \cite{RMP_Anatoli}, 
like the 
experiments on the real time evolution of essentially closed quantum systems in cold atomic gases \cite{Greiner,Kinoshita}. 
The experimental setup is typically a quantum quench, that is a sudden change of some parameter in the Hamiltonian. 
Therefore the system is initially prepared in a non-thermal superposition of the eigenstates of the Hamiltonian which 
drives its time evolution.

{}From a formal point of view, there is a very suggestive similarity between the canonical partition function of an
equilibrium system
\beq
Z(\beta)= \mbox{Tr}\, e^{-\beta H}
\label{def_Zbeta}
\eeq
and the overlap amplitude of some time-evolved initial quantum state $|\Psi_i\rangle$ with itself 
\beq
G(t)=\langle \Psi_i| e^{-iHt} |\Psi_i\rangle
\label{def_Gt}
\eeq
This leads to the question whether some analogue of temperature ($\beta$)-driven equilibrium phase
transitions in (\ref{def_Zbeta}) exists in real time evolution problems. In the theory of equilibrium phase
transitions it is well established that the breakdown of the high-temperature (small~$\beta$) expansion 
indicates a temperature-driven phase transition. Likewise, we propose the term 
{\em dynamical phase transition} for non-analytic behavior in time, that is the breakdown of a short time 
expansion in the thermodynamic limit at a critical time.
In this paper we study this notion of dynamical phase transition in the one dimensional transverse field Ising model, which
serves as a paradigm for one dimensional quantum phase transitions \cite{Sachdev}. It can be solved exactly,
which permits us to establish the existence of dynamical phase transitions that are intimately related
to the equilibrium quantum phase transition in this model. 

Our key quantity of interest is the boundary partition function
\beq
Z(z) = \langle \Psi_i|\,e^{-zH}\,|\Psi_i\rangle 
\label{defZ}
\eeq
in the complex plane $z\in\mathbb{C}$. For imaginary~$z=it$ this just describes the overlap
amplitude (\ref{def_Gt}). For real $z=R$ it can be interpreted as the partition function
of the field theory described by~$H$ with boundaries described by boundary states~$|\Psi_i\rangle$
separated by~$R$ \cite{LeClair}. In the thermodynamic
limit one defines the free energy density (apart from a different normalization)
\beq
f(z)=-\lim_{N\rightarrow\infty} \frac{1}{N}\,\ln\,Z(z)
\label{defF}
\eeq
where $N$ is the number of degrees of freedom. 
Now subject to a few technical conditions~\cite{MFisher} one can show that for finite $N$ the partition function (\ref{defZ}) 
is an entire function of~$z$ since inserting an eigenbasis of $H$ yields sums of terms
$e^{-zE_j}$, which are entire functions of~$z$. According to the Weierstrass factorization theorem \cite{Weierstrass}
an entire function with zeroes~$z_j\in \mathbb{C}$ can be written as
\beq
Z(z)=e^{h(z)}\prod_j \left(1-\frac{z}{z_j}\right)
\label{Zz}
\eeq
with an entire function $h(z)$. Thus
\beq
f(z)=-\lim_{N\rightarrow\infty} \frac{1}{N}\left[h(z)+\sum_j \ln\,\left(1-\frac{z}{z_j}\right)\right]
\eeq
and the non-analytic part of the free energy density is solely determined by the zeroes~$z_j$.
A similar observation was originally made by M.~E.~Fisher \cite{MFisher}, 
who pointed out that the partition function (\ref{def_Zbeta}) is an entire function in the complex temperature plane. 
This observation is analogous to the Lee-Yang analysis of equilibrium phase transitions in the complex magnetic field plane \cite{LeeYang}.
For example in the 2d~Ising model the Fisher zeroes in the complex
temperature plane approach the real axis at the critical temperature~$z=\beta_c$ in the thermodynamic limit,
indicating its phase transition~\cite{Saarloos}.

We now work out these analytic properties explicitly for the one dimensional transverse field Ising model
(with periodic boundary conditions)
\beq
H(g)=-\frac{1}{2}\sum_{i=1}^{N-1} \sigma_i^z\,\sigma_{i+1}^z+\frac{g}{2}\sum_{i=1}^N \sigma_i^x
\label{defH}
\eeq
For magnetic field $g<1$ the system is ferromagnetically ordered at zero temperature, and
a paramagnet for $g>1$ \cite{Sachdev}. These two phases are separated by a quantum critical point
at $g=g_c=1$. The Hamiltonian (\ref{defH}) can be mapped 
to a quadratic fermionic model \cite{LSM,Pfeuty,Barouch} 
\beq
H(g)=-\frac{1}{2}\sum_{i=1}^{N-1} \left(c^\dagger_i c_{i+1}+ c^\dagger_i c^\dagger_{i+1} +\mbox{h.c.}\right)
+ g \sum_{i=1}^{N} c^\dagger_i c_i
\label{Hf}
\eeq
Diagonalization yields the dispersion relation 
$\epsilon_k(g)=\sqrt{(g-\cos k)^2+\sin^2 k}$. 

In a quantum quench experiment the system is prepared in
the ground state for parameter~$g_0$, $|\Psi_i\rangle=|\Psi_{GS}(g_0)\rangle$, while
its time evolution is driven with a Hamiltonian~$H(g_1)$ with a different parameter~$g_1$.
In the sequel we will first analyze quench experiments in the setting of the fermionic model (\ref{Hf}).
A subtle difference occurs when thinking in terms of the spin model (\ref{defH}) since in the ferromagnetic
phase the ground state of the spin model is twofold degenerate, while the fermionic model always has
a unique ground state. We will say more about this later.
Taking the ground state of the fermionic model in Eq.~(\ref{Hf}) as the initial state $|\Psi_i\rangle$ the free energy density (\ref{defF})
describing this sudden quench $g_0\rightarrow g_1$  can be calculated analytically \cite{Silva} yielding
\beq
f_{g_0,g_1}(z) = -\int_0^\pi \frac{dk}{2\pi}\,\ln\left(\cos^2 \phi_k +\sin^2 \phi_k\,e^{-2z\epsilon_k(g_1)}\right)
\label{free_energy}
\eeq
Here $\phi_k=\theta_k(g_0)-\theta_k(g_1)$, and $\tan (2\theta_k(g))\stackrel{\rm def}{=} \sin k / (g-\cos k) \ , \ \theta_k(g)\in [0,\pi/2] \ .$
In (\ref{free_energy}) we have ignored an uninteresting additive contribution $z\,E_{GS}(g_1)/N$ that depends on the ground
state energy of~$H(g_1)$.% (in the notation of (\ref{Zz}) one has $h(z)=z\,E_{GS}(g_1)$). 

% For $g_0<1$ the ground state manifold of the Ising model is doubly degenerate implying that the initial state $|\Psi_i\rangle$ is not unique. In this 
% case we propose to generalize the return amplitude $G(t)$ in Eq.~(\ref{def_Gt}) to the Loschmidt matrix~\cite{Supp}. The experimentally relevant 
% quantity, however, is not $G(t)$ itself but rather the zero work limit of the work distribution function for a double quench $P(W=0,t)$, see below. 
% Although $G(t)$ depends on the precise choice of initial state, $P(W=0,t)$ does not~\cite{Supp}. This is of particular importance for the Ising 
% model where the ground state $|\Psi_{GS}(g_0)\rangle$ of the fermionic model in Eq.~(\ref{Hf}) is not the initial state for the quantum quench 
% protocol~\cite{Supp}. But still the return amplitude of $|\Psi_{GS}(g_0)\rangle$ determines completely $P(W=0,t)$~\cite{Supp}.% Note that these 
% subleties are absent for quenches starting in the paramagnetic phase.

\begin{center}
\begin{figure}
\includegraphics[width=8.5cm]{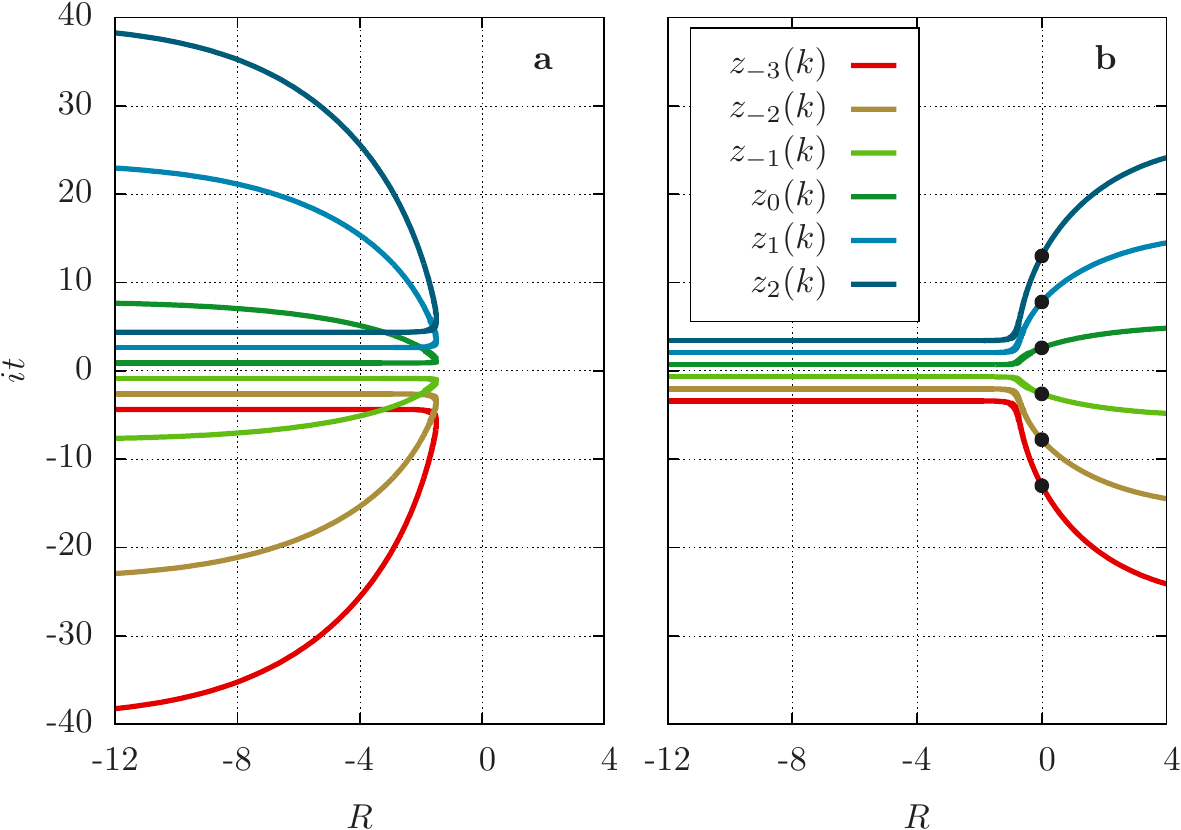}
\caption{Lines of Fisher zeroes for a quench within the same phase $g_0=0.4 \: \rightarrow \: g_1=0.8$ (left)
and across the quantum critical point $g_0=0.4 \: \rightarrow \: g_1=1.3$ (right). Notice that the Fisher
zeroes cut the time axis for the quench across the quantum critical point, giving rise to non-analytic
behavior at~$t_n^*$ (the times~$t_n^*$ are marked with dots in the plot).}
\label{fig_fisherzeroes}
\end{figure}
\end{center}

In the thermodynamic limit
the zeroes of the partition function in the complex plane coalesce to a family of lines labeled
by a number~$n\in\mathbb Z$
\beq
z_n(k)=\frac{1}{2\epsilon_k(g_1)}\,\left(\ln\tan^2 \phi_k+i\pi(2n+1)\right)
\eeq
The limiting infrared and ultraviolet behavior 
of the Boboliubov angles
\bea
\phi_{k=0}&=&\left\{ \begin{array}{ll} 0 & \mbox{quench in same phase} \\
\pi/4 & \mbox{quench to/from quantum critical point} \\
\pi/2 & \mbox{quench across quantum critical point} \end{array} \right. \nonumber \\
\phi_{k=\pi} &=& 0
\label{limit_phi}
\eea
immediately shows that the lines of Fisher zeroes cut the time axis for a quench across the quantum
critical point (Fig.~\ref{fig_fisherzeroes}) since then $\lim_{k\rightarrow 0} {\rm Re}\: z_n(k) = \infty$,
$\lim_{k\rightarrow\pi} {\rm Re}\: z_n(k) =- \infty$. 
In fact, the limiting behavior (\ref{limit_phi})
remains unchanged for general ramping protocols~\cite{Ramping}.

The free energy density (\ref{defF}) is just the rate function of the return amplitude 
%\beq
$G(t) = \exp[-N\,f(it)]$.
%\eeq
Likewise for the return probability (Loschmidt echo) 
$L(t)\stackrel{\rm def}{=} |G(t)|^2 = \exp(-N\,l(t))$
one has $l(t)=f(it)+f(-it)$. The behavior of the Fisher zeroes for quenches across the quantum
critical point therefore translates into non-analytic behavior of the rate functions for return
amplitude and probability at certain times~$t_n^*$. For sudden quenches one can work
out these times easily 
\beq
t^*_n=t^*\,\left(n+\frac{1}{2}\right) \ , \quad n=0,1,2,\ldots
\label{Fishertimes}
\eeq
with 
$t^*=\pi/\epsilon_{k^*}(g_1)$
and $k^*$ determined by $\cos k^*=(1+g_0\,g_1)/(g_0+g_1)$.
We conclude that for any quench across
the quantum critical point the short time expansion for the rate function of the return
amplitude and probability breaks down in the thermodynamic limit, analogous to the
breakdown of the high-temperature expansion at an equilibrium phase transition.
In fact, the non-analytic
behavior of $l(t)$ at the times~$t_n$ has already been derived by Pollmann et al.~\cite{Pollmann2009}
for slow ramping across the quantum critical point. For a slow ramping protocol $\epsilon_{k^*}(g_1)$  
becomes the mass gap $m(g_1)=|g_1-1|$
of the final Hamiltonian, but in general it is a new energy 
scale generated by the quench and depending on the ramping protocol. In the universal limit
for a quench across but very close to the quantum critical point, $g_1=1+\delta$, 
$|\delta|\ll 1$ and fixed $g_0$, one finds
$\epsilon_{k^*}(g_1)/m(g_1)\propto 1/\sqrt{|\delta |}$. Hence in this limit
the non-equilibrium energy scale $\epsilon_{k^*}$ becomes very different from the mass gap,
which is the only equilibrium energy scale of the final Hamiltonian. 

The interpretation of the mode $k^*$ follows from the observation $n(k^*)=1/2$, where 
$n(k)$ is the occupation of the excited state in the momentum $k$-mode in the eigenbasis of the final Hamiltonian $H_f(g_1)$. 
Modes $k>k^*$ have thermal occupation $n(k)<1/2$, while modes $k<k^*$ have inverted population
$n(k)>1/2$ and therefore formally negative effective temperature. The mode~$k^*$ corresponds
to infinite temperature. 
In fact, the existence of this infinite temperature mode 
and thus of the Fisher zeroes 
cutting the time axis periodically 
is guaranteed for arbitrary ramping protocols across the quantum critical point. For example, 
for slow ramping across the quantum critical point the existence of this mode and the negative temperature 
region in relation to spatial correlations was discussed in Ref.~\cite{Huse}.

\begin{center}
\begin{figure}
\includegraphics[width=8.5cm]{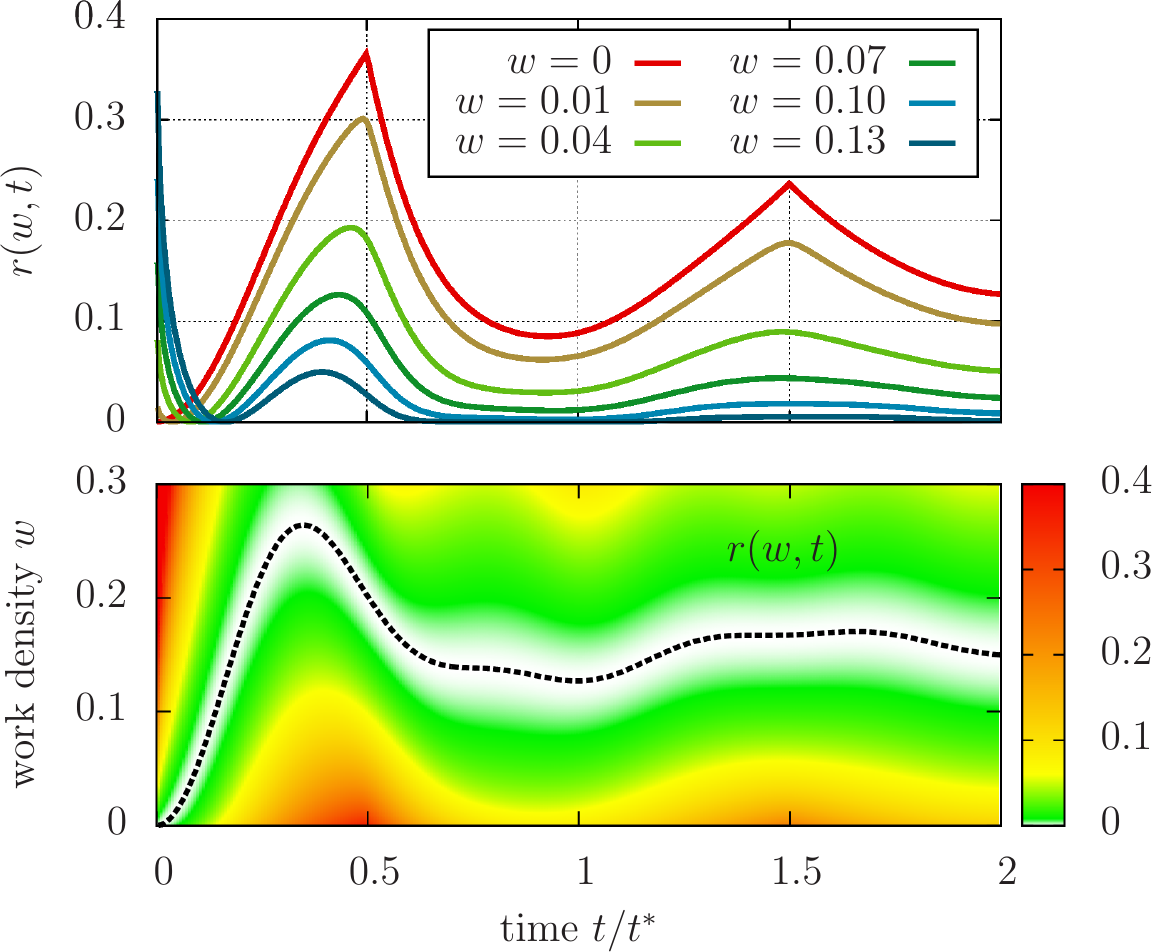}
\caption{The bottom plot shows the work 
distribution function $r(w,t)$ for a double quench across the quantum critical point
($g_0=0.5,\ g_1=2.0$). The dashed line depicts the expectation value
of the performed work, $r(w,t)=0$. The top plot shows various cuts for fixed values of the
work density~$w$. The line $w=0$ is just the Loschmidt echo: Its non-analytic
behavior at $t_n^*$ becomes smooth for~$w>0$, but traces of the non-analytic behavior
extend into the work density plane. In this respect work density plays a similar role to
temperature in the phase diagram of an equilibrium quantum phase transition.}
\label{fig_rwt}
\end{figure}
\end{center}
One measurable quantity in which the non-analytic behavior generated by the Fisher zeroes
appears naturally is the work distribution function of a double quench experiment: We prepare
the system in the ground state of $H(g_0)$, then quench to $H(g_1)$ at time $t=0$, and then
quench back to $H(g_0)$ at time~$t$. The amount of work~$W$ performed follows from the
distribution function
\beq
P(W,t)= \sum_j \delta\left(W-(E_j-E_{GS}(g_0))\right)\, |\langle E_j|\Psi_i(t)\rangle |^2
\label{doublequench}
\eeq
where the sum runs over all eigenstates $|E_j\rangle$ of the initial Hamiltonian~$H(g_0)$.
It obeys a large deviation form 
$
P(W,t)\sim e^{-N\,r(w,t)}
$
with a rate function $r(w,t)\geq 0$ depending on the work density $w=W/N$. In the thermodynamic
limit one can derive an exact result for $r(w,t)$: According to the G\"artner-Ellis theorem \cite{Touchette}
it is just the Legendre transform 
\beq
-r(w,t)=\inf_{R\in\mathbb{R}} \left(wR-c(R,t)\right)
\eeq
where 
\bea
c(R,t)&=&-\int_0^\pi \frac{dk}{2\pi} \ln\Big(
1 + \sin^2(2\phi_k) \sin^2( \epsilon_k(g_1) t ) \nonumber \\	
&&\qquad\qquad\qquad\qquad\times
( e^{-2 \epsilon_k(g_0) R}-1   )
\Big)
\eea
is the rate function for the cumulant generating function of the work distribution function,
$
C(R,t)=\int dW\,P(W,t)\,e^{-RW}=e^{-N\,c(R,t)}$.
In Fig. Fig.~\ref{fig_rwt} we show $r(w,t)$ for a quench
across the quantum critical point. For $w=0$ it
just gives the return probability to the ground state, $r(w=0,t)=l(t)$, therefore the
non-analytic behavior at the Fisher zeroes shows up as non-analytic behavior in the
work distribution function. However, from Fig.~\ref{fig_rwt} one can see that these
non-analyticities at $w=0$ also dominate the behavior for $w>0$ at~$t_n^*$, corresponding
to more likely values of the performed work. The suggestive similarity to the phase diagram
of a quantum critical point, with temperature being replaced by the work density~$w$, motivates
us to call this behavior dynamical {\it quantum} phase transitions. Notice that experimentally
the work density can be lowered by {\it post-selection} \cite{Supp}. 

So far we have analyzed the quench dynamics in terms of the fermionic model (\ref{Hf}). 
When thinking in terms of the transverse field Ising model (\ref{defH}), all results carry
over for quenches starting in the paramagnetic phase since then the spin ground state
is unique. Specifically, one finds the non-analytic behavior in the Loschmidt echo {\it and} the
work distribution function for quenches from the paramagnetic to the ferromagnetic phase.
For quenches originating in the ferromagnetic phase, the Loschmidt echo calculated above
corresponds to working in the Neveu-Schwarz sector \cite{Calabrese2012}, which amounts
to an unphysical superposition of spin up and spin down ground states in the spin language. 
However, looking at the experimentally relevant quantity work distribution function, 
one derives the same result  in the 
thermodynamic limit  as above when starting from either of the
two degenerate ferromagnetic ground states. Specifically, one obtains the non-analytic
behavior in $P(w=0,t)$ at the critical times (\ref{Fishertimes}) for quenches from the ferromagnetic
to the paramagnetic phase \cite{Supp}. 

\begin{center}
\begin{figure}
\includegraphics[width=8.5cm]{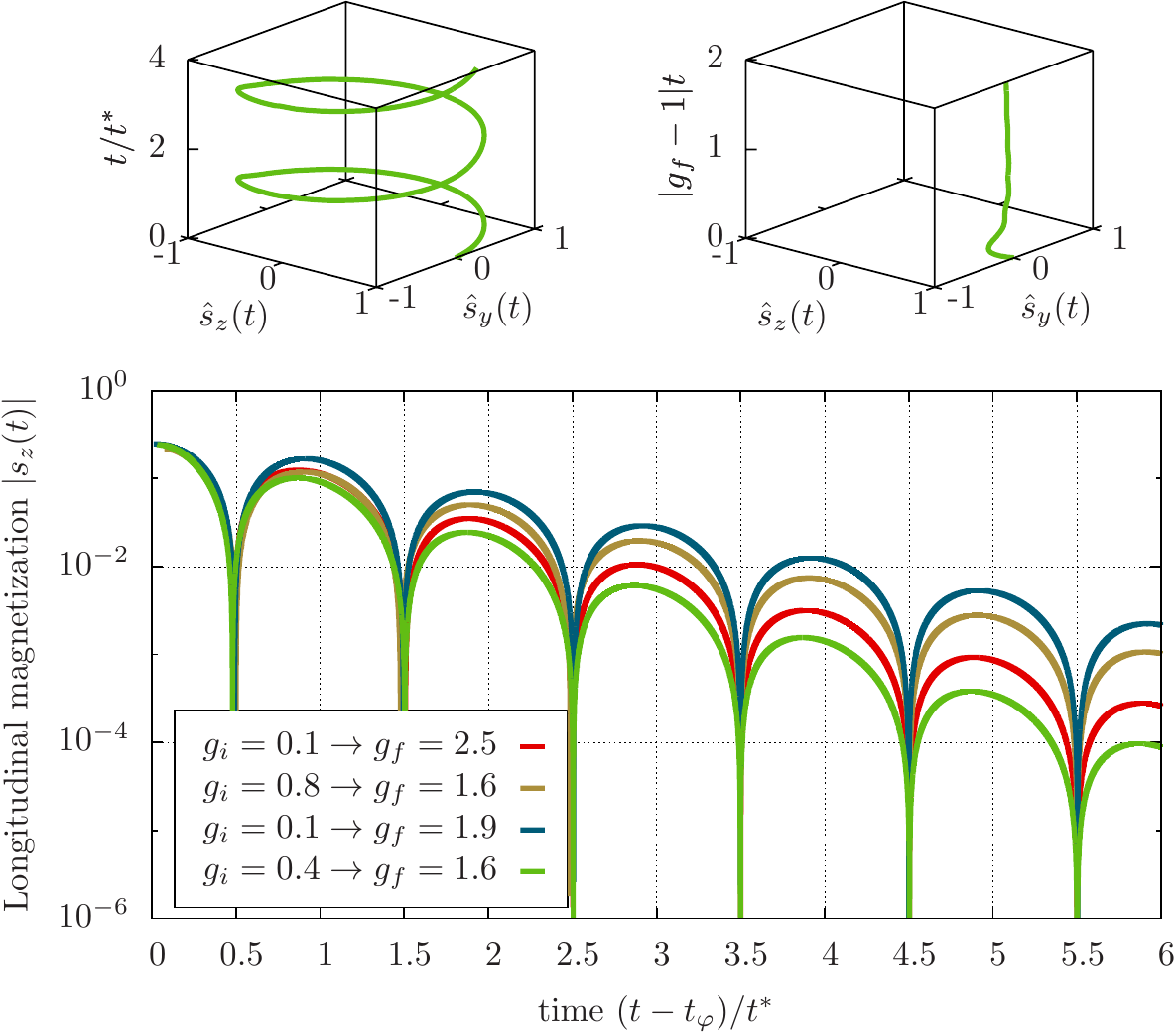}
\caption{Dynamics of the magnetization after the quench. The bottom plot shows the
longitudinal magnetization for various quenches across the quantum critical 
point. The time axis is shifted by a fit parameter~$t_\varphi$ and one can see that the
period of the oscillations is the time scale~$t^*$ (\ref{Fishertimes}). The upper plots show
the magnetization dynamics in the $y-z$-plane for a quench across the quantum
critical point $g_0=0.3 \: \rightarrow \: g_1=1.4$ (left) and a quench in the ordered phase 
$g_0=0.3 \: \rightarrow \: g_1=0.8$ (right). For better visibility the magnetization is 
normalized to unit length: $\hat s_{y,z}(t)\stackrel{\rm def}{=}s_{y,z}(t)/\sqrt{s_y^2(t)+s_z^2(t)}$.
Notice the Larmor precession
for the quench across the quantum critical point, while the dynamics for the quench
in the ordered phase is asymptotically just an exponential decay \cite{Calabrese2011}.}
\label{fig_orderparameter}
\end{figure}
\end{center}
Interestingly, the non-equilibrium time scale (\ref{Fishertimes}) also plays a role in
the dynamics of a local observable after the quench. We have calculated the longitudinal magnetization
by numerical evaluation of Pfaffians \cite{Barouch4}. 
For quenches within the ordered phase it is known analytically \cite{Calabrese2011,Schuricht} that
the order parameter decays exponentially as a function of time, which is expected
since in equilibrium one only finds long range order at zero temperature ($g<1$). For a quench
across the quantum critical point an additional oscillatory behavior is superimposed 
on this exponential decay, see Fig.~\ref{fig_orderparameter}. Notice that the behavior of the
magnetization remains perfectly analytic, but the
period of its oscillations agrees exactly (within numerical accuracy) with the period~$t^*$
of Fisher times. A conjecture consistent with our observation was also
formulated in Ref.~\cite{Calabrese2012}. A better understanding of this observation will be the topic of future work. At low energies the oscillatory decay transforms into real-time nonanalyticities at the Fisher times using the concept of post-selection allowing to observe the dynamical phase transitions in local observables~\cite{Supp}.

Summing up,
we have shown that ramping across the quantum critical point of the transverse field
Ising model generates periodic non-analytic behavior at certain times~$t_n^*$. This breakdown
of the short time expansion is reminiscent of the
breakdown of a high temperature expansion for the free energy at an equilibrium phase transition.
We have therefore denoted this behavior {\it dynamical phase transition}.
Notice that there are other related but not identical notions of dynamical phase
transitions, for example 
a sudden change of the dynamical behavior of an observable as  a function
of some control parameter \cite{Eckstein,Sciolla}, or qualitative changes in the ensemble of
trajectories as a function of the conjugate field of a dynamical order parameter \cite{Garrahan}.
% Our definition implies non-analytic behavior at some critical time, 
% which comes about due to the distribution of Fisher zeroes in the complex plane.

For quenches within the same phase (including to/from the quantum critical point)
the lines of Fisher zeroes lie in the negative half plane, ${\rm Re}\:z_j(k) \leq 0$ (Fig.~\ref{fig_fisherzeroes}).
Hence the knowledge of the equilibrium free energy $f(R)$ on the positive real axis completely
determines the time evolution by a simple Wick rotation.  This is no longer true for a quench/ramping protocol
across the quantum critical point since then the lines of Fisher zeroes cut the complex plane
into disconnected stripes, Fig.~\ref{fig_fisherzeroes}: Knowing $f(R)$ for $R\geq 0$ does not
determine the time evolution for $t>t_0^*$. In this sense non-equilibrium time evolution is no
longer described by equilibrium properties. 

The authors thank L.~D'Alessio, M.~Kolodrubetz and D.~Huse for valuable discussions. The authors also acknowledge the support of the Deutsche Forschungsgemeinschaft via SFB-TR~12, 
the German Excellence Initiative via the Nanosystems Initiative Munich (M.H. and S.K.),
the NSF under grants DMR-0907039, PHY11-25915, the AFOSR under grant FA9550-10-1-0110, the Sloan and Simons Foundations (A.P.). S.K. thanks the Boston University visitors program, A.P. and S.K. thank the Kavli Institute for Theoretical Physics at UCSB for their hospitality and NSF PHY11-25915.

\bibliographystyle{apsrev}

\appendix

\section{Loschmidt matrix}

For Hamiltonians with symmetry-broken ground states the definition of the return amplitude in Eq.~(\ref{def_Gt}) in the main text shares an ambiguity: there is not a unique ground state and therefore initial state $|\Psi_i\rangle$ for the nonequilibrium time evolution. Moreover, the return amplitude turns out to depend crucially on the precise choice of initial state in the degenerate ground state manifold for quenches across the quantum critical point in the Ising model for $g_0<1$ as will be shown in detail below. This is of particular importance because the ground state $|\Psi_{GS}(g_0)\rangle$ of the fermionic version of the Ising chain in Eq.~(\ref{Hf}) is not identical to one of the symmetry broken ferromagnetic ground states but rather to some superposition. The aim of this supplementary material is to demonstrate that, although the return amplitude depends on the precise choice of initial state, the experimentally relevant quantity, i.e., the work distribution of the double quench in Eq.~(\ref{doublequench}), does not. As a consequence, the free energy density of the fermionic model in Eq.~(\ref{free_energy}) is the relevant quantity although $|\Psi_{GS}(g_0)\rangle$ is not the initial state of the quantum quench protocol.

In case of symmetry broken ground states we propose the following generalization for the return amplitude that we will term the Loschmidt matrix:
\beq
	\mathcal{L}(t) =
	\left(
	\begin{array}{cc}\langle
	\Psi_1 | e^{-iH_ft} |\Psi_1 \rangle  & \langle \Psi_1 | e^{-iH_ft} |\Psi_2 \rangle \\
	\langle \Psi_2 | e^{-iH_ft} |\Psi_1 \rangle & \langle \Psi_2 | e^{-iH_ft} |\Psi_2 \rangle
	\end{array}
	\right)
\eeq
shown here for the case of two symmetry broken ground states $|\Psi_1\rangle$ and $|\Psi_2 \rangle$. The generalization to more than two states is straightforward. This matrix depends on the particular choice of the basis wave functions $|\Psi_1\rangle$ and $|\Psi_2\rangle$ in the degenerate ground state manifold. From a physical point of view, there is, however, a unique choice: the symmetry broken ground states. In case of vanishing transverse magnetic field in the Ising model this will be the fully polarized states $|\Psi_1\rangle=|\uparrow \dots \uparrow\rangle$ or $|\Psi_2\rangle = |\downarrow \dots \downarrow\rangle$, for example. 

Note that neither $|\Psi_{1}\rangle$ nor $|\Psi_{2}\rangle$ are the ground state $|\Psi_{GS}(g_0)\rangle$  of the fermionized model in Eq.~(\ref{Hf}) in the ferromagnetic phase. This can be seen directly by noticing that  $|\Psi_{GS}(g_0)\rangle$ carries no magnetization  $\langle \Psi_{GS}(g_0)|\sigma_l^z |\Psi_{GS}(g_0)\rangle=0$. Concluding $|\Psi_{GS}(g_0)\rangle$ rather is a superposition of $|\Psi_{1}\rangle$ and $|\Psi_2\rangle$. Fortunately, for most of the quantities considered in the literature such as the spin-spin correlation function $\sigma_l^z \sigma_m^z$ this makes no difference. This actually allows for the calculation of the absolute value of the magnetization even in the fermionic language using the cluster decomposition $\langle \Psi_{1/2}(g_0)|\sigma_l^z |\Psi_{1/2}(g_0)\rangle^2= \lim_{m\to\infty}\langle\Psi_{GS}(g_0)|\sigma_{l+m}^z \sigma_l^z |\Psi_{GS}(g_0)\rangle$~\cite{Barouch4}. Some quantities, however, such as the magnetization itself and the return amplitudes (as will be shown in detail below) the precise choice of the initial state can be crucial.

In the following we demonstrate the dependence of the return amplitude on the initial state for quenches from the ferromagnetic to the paramagnetic phase. We illustrate this subtle behavior for the particular example of a quench from $g_0=0$ to $g_1\gg1$. This case can be solved exactly and directly for the initial spin Hamiltonian in Eq.~(\ref{defH}) without using the mapping onto the fermionic language. For such a quench the Loschmidt matrix is given by
\beq
	\mathcal{L}(t) =
	\left(
	\begin{array}{cc}
	e^{-N f_{11}(t)}   &  e^{-N f_{12}(t)} \\
	e^{-N f_{21}(t)} & e^{-N f_{22}(t)}
	\end{array}
	\right)
\eeq
with
\bea
f_{11}(t) &=&f_{22}(t)= \log(|\cos(g_1t/2)|) \nonumber \\ f_{12}(t)&=&f_{21}(t) = \log(|\sin(g_1 t/2)|)
\eea
Here, we have ignored imaginary parts as they are not important for the discussion below. Note the large deviation scaling of the Loschmidt matrix elements that are exponentially suppressed with system size $N$. If $\mathrm{Re}[f_{11}(t)]>\mathrm{Re}[f_{12}(t)]$ the Loschmidt matrix is effectively diagonal in the thermodynamic limit $N\gg1$ whereas in the opposite case it only carries off-diagonal entries. For small times the diagonal elements typically dominate. But for sufficiently large times the off-diagonal components can build up indicating the possibility of transitions between the two different symmetry broken ground states. As we explain below this switching underlies the dynamical phase transition we are discussing.

The return amplitude $G(t)$ for the ground state $|\Psi_{GS}(g_0)\rangle$ of the fermionic model (which is a superposition of $|\Psi_{1}\rangle$ and $|\Psi_2\rangle$) and thus the free energy density $f(t)$ in Eq.~(\ref{free_energy}) is given by
\bea
	G(t) &=& e^{-N f(t)} = \langle \Psi_{GS}(g_0)|e^{-itH_F} |\Psi_{GS}(g_0)\rangle = \nonumber \\ & & = e^{-N f_{11}(t)} + e^{-N f_{12}(t)}.
\eea
Due to the large deviation scaling of the Loschmidt matrix elements the precise coefficients of the superposition are subleading and therefore do not appear in this equation. This is the case as long as no coefficient is exactly zero or scales exponentially with system size.
The diagonal elements of the free energy densities $f_{11}(t)=f_{22}(t)$ are analytic at the Fisher times $t_n^*=t^*(n+1/2)$ with $t^*=\pi/g_1$. Instead, they show logarithmic singularities at times $t=2t^*(n+1/2)\not=t_m^*$ for all $n,m\in \mathbb{Z}$. However, at the Fisher times $\mathrm{Re}[f_{11}(t_n^*)] = \mathrm{Re}[f_{12}(t_n^*)]$, i.e., $\sin(|g_1t_n^*/2|)=\cos(|g_1t_n^*/2|)$, and both the diagonal and off-diagonal contributions become identical. This signals a critical point where the dominant contribution to the Loschmidt matrix changes from the diagonal to the off-diagonal sector and vice versa. For the free energy density $f(t)$ this implies the following result
\beq
	f(t) = \left\{ \begin{array}{cl}
	               f_{11}(t) & \text{ , if } t_{2n-1}^* < t < t_{2n}^* \\
	               f_{12}(t) & \text{ , if } t_{2n}^* < t < t_{2n+1}^*
	               \end{array}
		\right.
		\, , \, n\in\mathbb{Z}.
\eeq
Concluding, the singularity in $f(t)$ is in fact not because the diagonal free energies are singular but rather because at the Fisher times $t_n^*$ the free energy switches between $f_{11}(t)$ and $f_{12}(t)$. Physically it implies that at the first Fisher time, for example, the return probability becomes dominated by the transition to a different magnetization sector.

Although the precise choice of initial state can be crucial for the return amplitude, in the following we demonstrate that the work distribution function $P(W,t)$ of a double quench defined in Eq.~(\ref{doublequench}) is independent of all these subleties. This is important as $P(W,t)$ is the experimentally measurable and therefore relevant quantity. In the zero work limit one obtains
\beq
	\lim_{W\to 0} P(W,t) = |\langle \Psi_1 | \Psi_0(t) \rangle|^2 + |\langle \Psi_2 | \Psi_0(t) \rangle|^2
\eeq
with $|\Psi_0(t)\rangle=e^{-iH_ft} | \Psi_0\rangle$ the time evolved initial state $|\Psi_0\rangle$. Irrespective of the precise choice of $|\Psi_0\rangle$ in the ground state manifold (could also be one of the symmetry broken ground states) $P(W\to0,t)$ always contains contributions from both magnetization sectors: 
\bea
	P(W\to0,t) &=& e^{-2N\mathrm{Re}[f_{11}(t)]} + e^{-2N\mathrm{Re}[f_{12}(t)]} = \nonumber \\
	& & =e^{-2N\mathrm{Re}[f(t)]}.
\eea
For the derivation of this equality the large deviation scaling of the probabilities is essentially important. Although $|\Psi_{GS}(g_0)\rangle$ is not the correct initial state for the quantum quench (it is not one of the symmetry broken ground states) its associated return amplitude with rate function $f(t)$  fully determines the experimentally relevant work distribution function:
\beq
	P(W\to 0 ) = \left|\langle \Psi_{GS}(g_0)|e^{-iH_f t}|\Psi_{GS}(g_0)\rangle\right|^2.
\eeq
The analysis in this supplementary material has illustrated the dependence of the return amplitude on the choice of initial state for the extreme quench from $g_0=0$ to $g_1\gg1$. Although the return amplitude depends on this choice, the work distribution for a double quench does not.

It is possible to also include $1/g_1$ corrections in the calculation of the return amplitude using the original spin language for large but finite magnetic fields $g_1$. It turns out that this improved treatment fully supports the conclusions drawn above. This analysis, however, is beyond the scope of this supplementary material and will be presented elsewhere.

\section{Postselection of observables}

One route towards the experimental observation of the dynamical phase transitions in local observables is the idea of post-selection.
Let us consider the double-quench experiment as discussed in the main text but with a slightly additional twist. Namely we assume that the system is prepared in the ground state of the initial Hamiltonian (actually the initial state can be arbitrary) then the Hamiltonian of the system is quenched to the final Hamiltonian $H_f$ for the time $t$ and then quenched back to the initial Hamiltonian $H_i$. After that we allow the system to relax to the diagonal ensemble~\cite{Rigol2008}, which using a different language also means measuring the energy of the system~\cite{RMP_Anatoli}. In the resulting diagonal state we measure the expectation value of an arbitrary observable $\mathcal O$ as a function of $H_f$ and $t$. Note that for observables commuting with $H_i$ like energy the projection to the diagonal ensemble is not affecting the result, while for the observables which do not commute with $H_i$ such projection makes a difference.

The expectation value of  the observable $\mathcal O$ is then
\beq
\langle \mathcal O(t)\rangle=\sum_n |\langle \psi(t)|n\rangle|^2 \langle n|\mathcal O|n\rangle=\sum_n p_n(t) \langle n|\mathcal O|n\rangle,
\eeq
where $|n\rangle$ denotes the energy eigenstates of $H_i$ and $p_n(t)= |\langle \psi(t)|n\rangle|^2$ are the probabilities of occupation of these states.
Within a continuum description the sum above can be formally written as a continuous integral over energies
\beq
\langle \mathcal O(t)\rangle=\int dE P(E,t) {\mathcal O(E,t)}, 
\eeq
where $P(E,t)=\sum_n p_n(t) \delta (E-E_n)$ is the work distribution function for a double quench and
\beq
	\mathcal{O}(E,t) = \sum_n \frac{p_n(t)}{P(E,t)} \langle n | \mathcal O |n \rangle \delta (E-E_n).
\eeq
\begin{figure}
\centering
\includegraphics[width=0.9\linewidth]{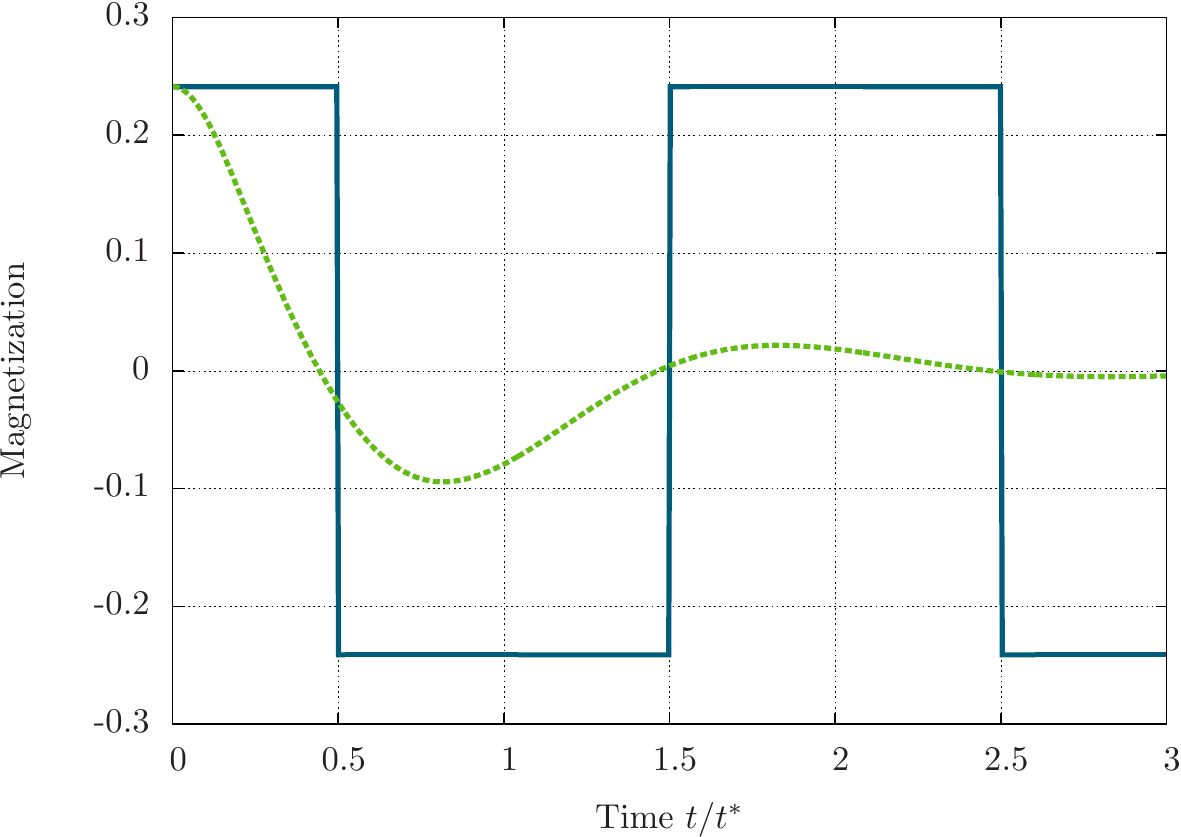}
\caption{(Color online) Dynamics of the post-selected magnetization after a quench from $g_i=0.5$ to $g_f=1.5$. While the blue (solid) curve shows the order parameter dynamics after a quench, the green (dashed) curve shows the zero-temperature limit $\tilde \beta \to \infty$ of the post-selected magnetization.}
\label{fig_top}
\end{figure}

{\em Postselection.} The idea of postselection is that one can artificially skew the energy distribution $P(E,t)$ by for example disregarding instances with energy above the certain threshold. From computational purposes it is more convenient, however, to skew the distribution by multiplying it by an exponential factor $\exp(-\tilde \beta E)$, where $\tilde \beta$ plays the role of postselected temperature. We thus can formally define
\beq
P_{\tilde \beta} (E,t) =\frac{1}{\tilde Z_{\tilde \beta}(t)} P(E,t) e^{-\tilde \beta E},
\eeq
where $\tilde Z_{\tilde \beta}(t)$ is the postselected partition function:
\beq
\tilde Z_{\tilde\beta}(t)=\int dE P(E,t) e^{-\tilde\beta E}=\sum_n p_{\tilde \beta,n}(t),
\eeq
with $p_{\tilde \beta,n}(t) = p_n(t) \exp[-\tilde\beta E]$. Then the post-selected expectation value of the observable will read:
\beq
\langle \mathcal O(t)\rangle_{\tilde\beta}=\sum_n  p_{\tilde \beta,n} \langle n | {\mathcal O} | n \rangle.
\eeq
yielding in the continuum description:
\beq
\langle \mathcal O(t)\rangle_{\tilde\beta}=\int dE \tilde P_{\tilde \beta} (E,t) {\mathcal O_{\tilde \beta}(E,t)}.
\eeq
We will not analyze in detail this postselection procedure here and its similarities and differences with thermodynamics (which is equivalent to preselection in our language) since this lies beyond the scope of our work. We only note that as the postselected temperature $\tilde T=\tilde \beta^{-1}$ is lowered we are effectively projecting the observable to the ground state manifold. In Fig.~\ref{fig_top} we show the postselected magnetization in this low temperature limit reavealing jumps between the different symmetry broken sectors, which are located precisely at the Fisher times. Note that this allows one to observe real-time nonanalyticities in local observables. We anticipate that in this limit the magnetization dynamics will be decoherence free at the expence of effectively excluding large amounts of data from the analysis. This point will be a subject of our future work.

\section{Postselection and the complex time return amplitude}

Let us make another brief point where postselection can be used to obtain nontrivial results. We now go back to a single quench of the Hamiltonian from $H_i$ to $H_f$. Then as it was shown in Ref.~\cite{Silva} up to the phase factor the return amplitude $G(t)$ is the Fourier transform of the work distribution after the quench:
\beq
G(t)=\int dW P(W) \exp[i Wt],
\eeq
where 
\beq
P(W)=\sum_n |\langle \psi_i|n^f\rangle|^2 \delta (E^f_n-E_0-W),
\eeq
where $|n^f\rangle$ are the energy eigenstates of the final Hamiltonian and $E^f_n$ are the corresponding eigenenergies.

Using similar considerations as in the previous section we can define the postselected work distribution function (but this time with respect to the eigenstates of the final Hamiltonian):
\beq
\tilde P_{\tilde \beta} (W)=\frac{1}{\tilde Z_{\tilde \beta}} P(W)\exp[-\tilde \beta W].
\eeq 
Then trivially extending the analysis of Ref.~\cite{Silva} we find that the inverse Fourier transform of the postselected work distribution function gives the return amplitude at a complex time:
\beq
\int dW \tilde P_{\tilde \beta}(W) \exp[i Wt]=G(t+i\tilde \beta)
\eeq
Thus amazingly the complex time return amplitude and hence the Fisher zeros analyzed in Fig. 1 of the main paper are measurable quantities at least in principle (note that the postselected temperature can be both positive and negative).

\bibliographystyle{apsrev}

\begin{thebibliography}{30}
\bibitem{Ramping}

For a general ramping protocol $g(t)$ with $g(t=0)=g_0, g(t=\tau)=g_1$ we define
$|\Psi_i\rangle=|\psi(\tau)\rangle$ and $|\psi(t)\rangle$ is the solution
of the Schroedinger equation: $i\,\partial_t\,|\psi(t)\rangle = H(g(t))\,|\psi(t)\rangle \ , \, |\psi(t=0)\rangle=|\Psi_{GS}(g_0)\rangle
$
\end{thebibliography}

\end{document}